\documentclass[letterpaper,prb,nobalancelastpage,twocolumn]{revtex4}
\usepackage{amsmath,amssymb,graphicx,times,setspace}

\begin{document}
\title{Model for the free-volume distributions of equilibrium fluids}
\author{William P. Krekelberg} 
\email{krekel@che.utexas.edu} 
\author{Venkat Ganesan}
\email{venkat@che.utexas.edu} 
\thanks{Alfred P. Sloan Fellow}
\author{Thomas M. Truskett} 
\email{truskett@che.utexas.edu}
\thanks{Corresponding author} 
\affiliation{Department of Chemical Engineering and Institute for
  Theoretical Chemistry, The University of Texas at Austin, Austin, TX 78712.}

\newcommand{\subf}[1]{{#1}_{\mathrm f}}
\newcommand{\subfs}[1]{{#1}_{\mathrm f}^*} 
\def\xf{ \subf{x} }
\def\xfs{ \xf^*} 
\def\pgap{{p_{\mathrm g} } } 
\def\GS{G\"{u}rsey} 
\def\vf{ \subf{v} }
\def\sf{ \subf{s} }
\def\vfs{ \vf^* } 
\def\sfs{ \sf^* }
\def\Avf{{\langle \vf \rangle}} 
\def\Asf{{\langle \sf \rangle}}
\def\Axf{{\langle \xf \rangle}}
\def\pvfo{f_{\mathrm{1D}}}
\def\pvf { f }
\def\psf { \phi }
\def\pvfs { f^* }
\def\psfs { \phi^* }
\def\ns{\mspace{-15mu}}
\def\intI{\int_0^\infty\ns}
\begin{abstract}
  We introduce and test via molecular simulation a simple model for
  predicting the manner in which interparticle interactions and
  thermodynamic conditions impact the single-particle free-volume
  distributions of equilibrium fluids. The model suggests a scaling
  relationship for the density-dependent behavior of the hard-sphere
  system.  It also predicts how the second virial coefficients of
  fluids with short-range attractions affect their free-volume
  distributions.
\end{abstract}
\maketitle

\section{Introduction}

Liquid-state theory aims to provide a framework that links the
interparticle interactions of a fluid with its local structure,
thermodynamic properties, and transport coefficients.  One of the
central quantities for characterizing the structural order is the
static structure factor $S(k)$ (or, equivalently, the pair correlation
function $g(r)$).\cite{Hanson1986Theory-of} The structure factor can
be readily measured by scattering experiments, computed via molecular
simulations, or estimated using integral equation theories.
Thermodynamic properties of fluids with pairwise interactions can be
calculated directly from $S(k)$ using exact relationships from
statistical mechanics.  Furthermore, many of the nontrivial dynamical
behaviors of liquids can be predicted from a knowledge of $S(k)$ using
mode-coupling theory and its recent
extensions.\cite{Hanson1986Theory-of,Gotze1999,Reichman2005mode}

However, despite its considerable practical value, $S(k)$ cannot
provide a comprehensive description of liquid structure mainly because
it only contains information about the spatial correlations between
pairs of particles.  Higher-order correlation functions, or suitable
approximations for them, are required to predict structural quantities
that depend on the relative positions of three or more particles.  A
well known example of such a quantity is the single-particle free
volume $\vf$, illustrated in Fig.~\ref{fig:schematic_fvmodel} .  It is
defined as the cage of accessible volume that a given particle center
could reach from its present state if its neighboring particles were
fixed in their current configuration.\cite{Hoover1972Exact-Dynamical}
Restated in simple terms, $\vf$ quantifies the ``breathing room'' that
a particle has in its local packing environment.  The thermodynamic
properties of purely athermal (i.e., hard-core) fluids can be formally
related to the statistical geometry of their single-particle free
volumes.\cite{Hoover1972Exact-Dynamical,Speedy1981Cavities-and-fr,Speedy1991,Sastry1998Free-volume-in-,Corti1999Statistical-geo}

The idea that relaxation processes should also be linked to free
volumes has a long history in studies of the liquid
state,\cite{A.1966Liquid-Structur,Cohen1959Molecular-Trans,Cohen1979}
and it has motivated the development of numerous ``free-volume based''
models for predicting transport
coefficients.\cite{Liu2002Genaralised-fre} Unfortunately, a general
microscopic framework has yet to emerge, in part due to the
computational and experimental difficulties associated with measuring
and characterizing free volumes.  However, recent advances in
computational statistical
geometry\cite{Sastry1998Free-volume-in-,Sastry1997statistical-geo}
have made it possible to efficiently calculate such properties from
particle configurations obtained via either experiments (e.g.,
confocal microscopy of colloidal
suspensions\cite{Conrad2005Weak-Correlatio,Dullens2006Direct-measurem})
or computer
simulations.\cite{Sastry1998Free-volume-in-,Truskett1999The-Statistical,Starr2002What-do-we-lear,krekelberg2006}
These methods have catalyzed new efforts to quantitatively examine the
basic ideas underlying the free-volume perspective for dynamics and,
hence, the prospects for developing a successful microscopic theory.

In this article, we contribute to one aspect of this effort by
introducing a simple model for predicting the statistical geometry of
single-particle free volumes in equilibrium fluids.  We use the
aforementioned computational tools of statistical geometry to test the
predictions of the model for (i) the hard-sphere (HS) fluid and (ii) a
fluid of particles with short-ranged, square-well attractions.  The
former is the standard structural reference fluid for simple liquids.
The latter serves as an elementary model for the behavior of
suspensions of attractive colloids and globular proteins, whose
structures can be strongly influenced by interparticle attractions.
As we show, our model suggests a scaling relationship for the
density-dependent behavior of the HS system.  It also predicts the
manner in which the second virial coefficients of fluids with
short-range attractions affect their free-volume distributions.

\section{General Framework}  
We consider a three-dimensional (3D) equilibrium fluid comprising $N$
identical spherical particles contained in a macroscopic volume $V$ at
temperature $T$.  The packing fraction is $\eta=N \pi \sigma^3 /6V$,
and the particles interact via a short-range isotropic pair potential
$V_{ij}(r)$ of the generic form
\begin{equation}
  \label{eq:vij-gen}
  V_{ij}(r)=
  \begin{cases}
    \infty&r<\sigma,\\
    u(r)&\sigma\leq r < \sigma(1+\Delta),\\
    0&r\geq \sigma(1+\Delta),
  \end{cases}
\end{equation}
where $\Delta \le 1$.  If one chooses $u(r)=-\epsilon$, then
Eq.~\eqref{eq:vij-gen} describes a square-well interaction.
Alternatively, the hard-sphere potential is recovered if $\Delta=0$.
We are interested in developing a general strategy for predicting how
$V_{ij}(r)$, $\eta$, and $T$ affect the statistical properties of the
fluid's single-particle free volumes.  To simplify notation, we
implicitly non-dimensionalize all lengths from this point forward by
the hard-core diameter $\sigma$ of the particles.

\begin{figure}[t]
  \centering
  \includegraphics{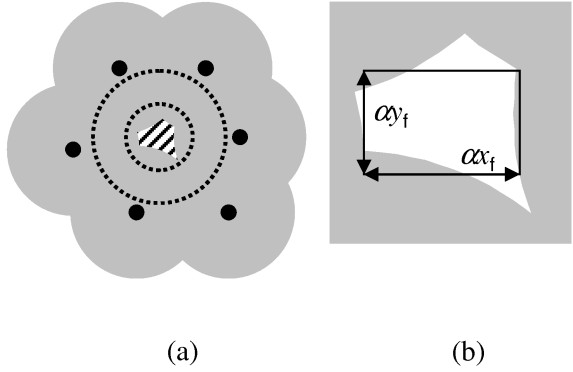}
  \caption{A 2D schematic of the free volume of a tagged
    particle.  Overlapping grey circles represent the exclusion spheres of the
  neighboring particles.  (a) The small dashed circle is the tagged 
particle surface,
  and the larger dashed
  circle is its associated exclusion sphere.  The cross-hatched region is
  the tagged particle's free volume.  (b) An expanded view of the tagged
  particle's free volume along with its approximate representation
  in our model.}
  \label{fig:schematic_fvmodel}
\end{figure}

The problem described above cannot be treated exactly, and so we make
some simplifying approximations that we later assess by comparing our
predictions to results obtained via molecular simulations.  Our basic
working assumption is that the statistical geometry of the free
volumes in the 3D fluid at $T$ and $\eta$ can be predicted based on a
knowledge of the exact free-volume distribution $f_{1{\mathrm D}}$ of
the corresponding one-dimensional (1D) fluid at the same temperature
$T$ and the scaled packing fraction $\xi = \eta/\eta_{\mathrm{MRJ}}$.
We choose $\eta_{\mathrm{MRJ}}=0.64$ to ensure that the close packing
limit of the 1D fluid ($\xi=1$) maps onto the maximally random jammed
state of the 3D system.\cite{Torquato2000} Our specific idea is that a
particle's free volume in the 3D fluid at $T$ and $\eta$ can be
modeled as a cuboid (see Fig.~\ref{fig:schematic_fvmodel}) with length
$\alpha \subf{x}$, height $\alpha \subf{y}$, and width $\alpha
\subf{z}$, where $\xf$, $\subf{y}$, and $\subf{z}$ are independent
random variables drawn from $f_{1{\mathrm D}}$ at $T$ and $\xi$.  
The 3D free-volume distribution $f(\vf)$ is thus
\begin{multline} 
  \label{eq:3Dify}
  f(\vf)= \intI dx\! \intI dy\! \intI dz\ \delta(\vf-\alpha^3xyz)\\
 \times \pvfo (x) \pvfo (y) \pvfo (z). 
\end{multline} 
The constant $\alpha$ is a scale factor chosen to ensure that our
construction accurately reproduces the equilibrium free-volume
distributions and thermodynamic properties of the 3D HS fluid
(discussed below).  Also of interest is the probability density
$\phi(\sf)$ associated with finding a 3D free volume with surface area
$\sf$.  Using our model, $\phi(\sf)$ can be expressed as
\begin{multline}
  \label{eq:fsf}
  \phi(\sf)=\intI dx\! \intI dy\! \intI dz\ \delta(\sf-2\alpha^2[xy+xz+yz])\\
\times \pvfo(x) \pvfo(y) \pvfo(z).
\end{multline}
Eq.~\eqref{eq:3Dify} and \eqref{eq:fsf} indicate that, in order to predict $f(\vf)$ and $\phi(\sf)$ using our approach, 
one only needs to
derive the state-dependent form of $\pvfo$ from knowledge of
$V_{ij}(r)$.  However, $\pvfo$ is given by
\begin{multline}
  \label{eq:pvf_1D_gen}
  f_{1 {\mathrm{D}}}(\xf) =\intI dy_L\! \intI dy_R\ \delta(\xf
   -y_L-y_R) \\
   \times
\pgap(y_L)\pgap(y_R)dy_L dy_R,
 \end{multline}
 where $\pgap(z)$ is the probability density associated with finding a
 gap of size $z$ between the surfaces of two neighboring particles in
 the corresponding 1D fluid.  For particles that interact via 
a pair potential of the form given by Eq.~\eqref{eq:vij-gen},
 \GS\cite{Gursey1950Classical-stati} has shown that $\pgap(z)$ can be written
\begin{equation}
  \label{eq:gap_def}
  \pgap(z)=\frac{e^{-\beta \Pi [z+1]}e^{-\beta V_{ij}(z+1) }
  }{\int_0^\infty e^{-\beta \Pi [s+1]}e^{-\beta V_{ij}(s+1)} ds}, 
\end{equation}
where $\beta=(k_{\mathrm B} T)^{-1}$, and $\Pi$ is the 1D pressure.
The equation of state of the 1D fluid, i.e., the dependence of $\Pi$ on 
$\beta$ and $\xi$, can be found implicitly from the following
relation:\cite{Gursey1950Classical-stati}
\begin{equation}
  \label{eq:eos1d-1}
  \beta \xi^{-1}=-\left(\frac{\partial \ln \psi}{\partial \Pi}\right)_{T},
\end{equation}
where
\begin{equation}
  \label{eq:eos1d-2}
  \psi=\intI dx\ e^{-\beta \Pi x}e^{-\beta V_{ij}(x)}.
\end{equation}

Having outlined the general strategy of our model, we examine some of
its predictions for two specific systems in Section III: (i) the
equilibrium HS fluid and (ii) an equilibrium square-well fluid with
short-range attractions.

\section{Testing the model}
\subsection{Predictions for the HS fluid} 
Here we apply the model of Section II to predict
the statistical geometry of the free volumes in the equilibrium HS
fluid, where the pair potential $V_{ij}(r)$ is given by 
\begin{equation}
  \label{eq:vij-HS}
  V_{ij}(r)=
  \begin{cases}
    \infty&r<1,\\
    0&r\geq 1.
  \end{cases}
\end{equation}
From Eq.~\eqref{eq:vij-HS}, \eqref{eq:eos1d-1}, and
\eqref{eq:eos1d-2}, it follows that the equation of state of the 1D HS
fluid is given by\cite{Tonks1936}
\begin{equation}
\label{eq:1d-hs-eos}
\beta \Pi/\xi=1/(1-\xi).  
\end{equation}
Substituting Eq.~\eqref{eq:1d-hs-eos} and \eqref{eq:vij-HS} into 
Eq.~\eqref{eq:gap_def}, yields the corresponding gap-size distribution,
\begin{equation}
  \label{eq:pgap_1D_hs}
  \pgap(z)=\frac{\xi}{1-\xi}e^{-z \xi /(1-\xi)}.
\end{equation}
Then, from Eq.~\eqref{eq:pvf_1D_gen}, $\pvfo(\xf)$ is simply
\begin{equation}
  \label{eq:pvf_1D_hs}
    \pvfo(\xf) =\xf\left( \frac{\xi}{1-\xi} \right)^2 e^{-\xf\xi/(1-\xi)}.
\end{equation}
The first moment of this distribution is
\begin{equation}
  \label{eq:ave_vf_1D}
  \Axf\equiv \intI d\xf\ \xf \pvfo (\xf) =\frac{2(1-\xi)}{\xi}.
\end{equation}
Defining $\xf^* \equiv \xf/\Axf$, we have
\begin{equation}
  \label{eq:pvf_1D_relation}
  \pvfo(\xf)d\xf=\pvfo^*(\xf^*)d\xf^*,
\end{equation}
where
\begin{equation}
  \label{eq:pvf_universal_1D_hs}
  \pvfo^*(\xf^*)=4 \xf^* e^{-2 \xf^*}.
\end{equation}
The main implication is that while $\pvfo$ is a function of $\xf$ and
the packing fraction $\xi$ for the 1D HS fluid, $\pvfo^*(\xf^*)$ can
be represented by a single curve that is independent of $\xi$.  As we
show below, this property, when used in combination with
Eq.~\eqref{eq:3Dify} and \eqref{eq:fsf} of our model, leads to scaling
predictions for the density-dependent free-volume and free-surface
distributions of the 3D HS fluid.

In particular, making use of
Eq.~\eqref{eq:3Dify}, \eqref{eq:pvf_1D_hs}, and \eqref{eq:ave_vf_1D}, 
we identify that
\begin{equation}
  \label{eq:ave_vf_3D}
  \Avf \equiv \intI dv\ v \pvf(v) =\alpha^3 \Axf^3=8\alpha^3\left( \frac{1-\xi}{\xi} \right)^3.
\end{equation}
Similarly, combining Eq.~\eqref{eq:fsf}, \eqref{eq:pvf_1D_hs}, and 
\eqref{eq:ave_vf_1D} yields
\begin{equation}
  \label{eq:ave_sf_3D}
  \Asf\equiv \intI ds\ s\psf(s) =6 \alpha^{2} \Axf^2=24\alpha^{2}\left(
    \frac{1-\xi}{\xi} \right)^2.
\end{equation}
Now, if we normalize the 3D free volume by its first moment,
$\vfs \equiv \vf/\Avf$, then we have
\begin{equation}
  \label{eq:3D_fv_scaled_relation}
  \pvf(\vf) d\vf = \pvfs(\vfs)d\vfs,
\end{equation}
where
\begin{multline}
  \label{eq:3Dify_scaled}
  \pvfs(\vfs)=\intI dx\! \intI dy\! \intI dz\ \delta(\vfs- xyz)\\
\times
\pvfo^*(x) \pvfo^*(y) \pvfo^*(z).
\end{multline}
Similarly, if we introduce the reduced free surface $\sfs\equiv \sf/\Asf$,
then it follows that 
\begin{equation}
  \label{eq:3D_fs_scaled_relation}
  \psf(\sf) d\sf=\psfs(\sfs)d\sfs,
\end{equation}
where
\begin{multline}
  \label{eq:3Dify_scaled_fs}
  \psfs(\sfs)=\intI dx\! \intI dy\! \intI dz\ \delta(\sfs-[xy+xz+yz]/3
  )\\
\times
\pvfo^*(x) \pvfo^*(y) \pvfo^*(z) .
\end{multline}
Eq.~\eqref{eq:3Dify_scaled} and~\eqref{eq:3Dify_scaled_fs}, when
viewed together with Eq.~\eqref{eq:pvf_universal_1D_hs}, show that our
model predicts that the scaled free-volume and free-surface
distributions, $\pvfs(\vfs)$ and $\psfs(\sfs)$, are independent of
packing fraction for the 3D HS fluid.  We will return to this point
later, when we test the predictions of the model via molecular
simulations.

To use the model to predict the shapes of the free volumes in the 3D
HS fluid, we analyze the behavior of the dimensionless sphericity
parameter $\lambda_{\mathrm f}$, defined
as\cite{Ruocco1991,Ruocco1992}
\begin{equation}
  \label{eq:def_sphere}
  \lambda_{\mathrm{f}} = \sf (6 \pi^{1/2} \vf)^{-2/3}.
\end{equation}
The sphericity parameter takes on its minimum possible value
($\lambda_{\mathrm f} = 1$) for a spherical free volume, while 
it is larger in magnitude
for less symmetric free volumes.  
Our model predicts that the average value of 
the sphericity parameter (quantifying the average shape of the free
volumes), 
given by
\begin{multline}
  \label{eq:ave_sphereicity}
  \langle \lambda_{\mathrm{f}} \rangle =\intI dx\! \intI dy\! \intI
  dz\ \frac{2(xy+xz+yz)}{(6\pi^{1/2}xyz)^{2/3}} \\
\times\pvfo(x) \pvfo(y) \pvfo(z)\simeq 1.57,
\end{multline}
is independent of packing fraction for the 3D HS fluid.

Finally, the
equation of state of the 3D HS fluid 
can be formally related to the statistical
geometry of its free volumes:\cite{Hoover1972Exact-Dynamical,Speedy1981Cavities-and-fr,Sastry1998Free-volume-in-}
\begin{equation}
  \label{eq:eos}
  \frac{\beta P}{\rho}=1+\frac{1}{6}\left< \frac{\sf}{\vf} \right>.
\end{equation}
Here, $P$ is the pressure of the 3D fluid, and $\rho=N/V$ is its 
number density.  In our model, it is easily shown that
\begin{multline}
  \label{eq:ave_sf_o_vf}
  \left<\frac{\sf}{\vf}\right>=\intI dx\!\intI dy\! \intI dz\
  \frac{2(xy+xz+yz)}{\alpha xyz}\\
\times \pvfo(x) \pvfo(y) \pvfo(z) =\frac{6\alpha^{-1}\xi}{1-\xi} ,
\end{multline}
and therefore we have 
\begin{equation}
  \label{eq:eos_sf_vf}
  \frac{\beta P}{\rho}=\frac{1-(1-\alpha^{-1})\xi}{1-\xi}.
\end{equation}

\subsection{Simulations of the HS Fluid}
To test the predictions of our free-volume model for the 3D
equilibrium HS fluid, we have performed a series of molecular dynamics
simulations using a standard event-driven
algorithm.\cite{Rapaport2004The-Art-of-Mole} All runs were carried out
in the microcanonical ensemble using $N=1000$ particles and a
periodically-replicated cubic simulation cell.  Snapshots of the
system's equilibrium configurations were collected and used to
calculate the geometric properties of single-particle free volumes via
the exact algorithm presented by Sastry \textit{et
  al.}.\cite{Sastry1998Free-volume-in-,Sastry1997statistical-geo}

\begin{figure}[t]
  \centering
  \includegraphics{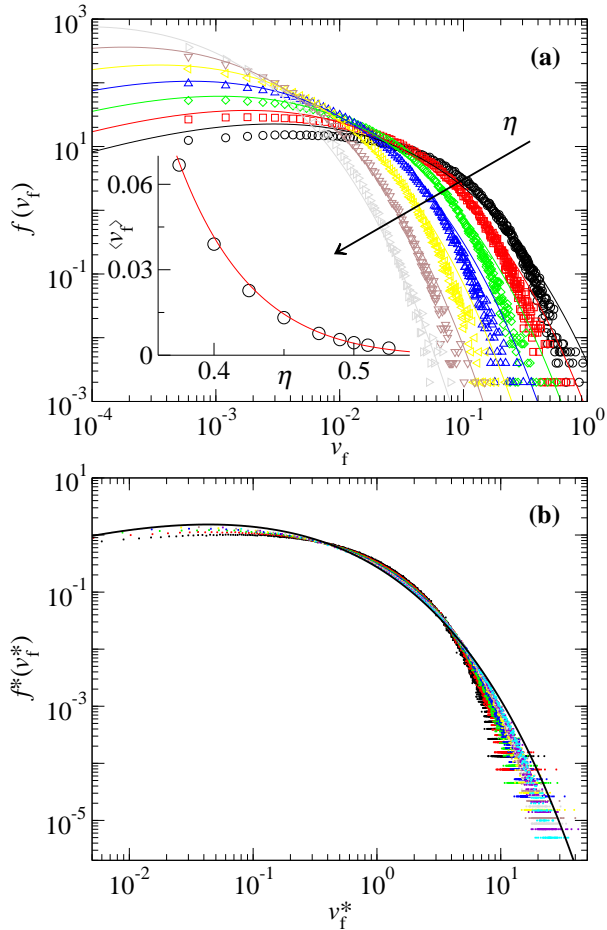}
  \caption{(a) Free-volume distributions of the 3D HS fluid for
    packing fractions $\eta=0.375,\ 0.4,\ 0.425,\ 0.45,\ 0.475,\ 0.5,$
    and $0.525$.  Symbols are simulation data and solid curves are
    the predictions of Eq.~\eqref{eq:3Dify}.  
The arrow indicates increasing packing
  fraction~$\eta$. Inset: Average free volume $\Avf$
    versus packing fraction $\eta$. Circles are simulation data and
    the solid line is the prediction
    of Eq.~\eqref{eq:ave_vf_3D}.  (b) Probability density associated
    with observing particles with scaled free volume 
$\vf^*=\vf/\Avf$.  Symbols are
    simulation data from panel~(a) and the solid line
    is the prediction of
    Eq.~\eqref{eq:3Dify_scaled}.}
  \label{fig:HS-fv}
  
\end{figure}

Numerical predictions using the free-volume model introduced here
require specifying the value of the geometric scale factor~$\alpha$
shown in Fig.~\ref{fig:schematic_fvmodel}.  As discussed in Section
II, the goal is to choose $\alpha$ in a way that allows the model to
provide a reasonable overall description of the statistical geometry,
and hence the thermodynamic properties, of the HS fluid.  Throughout
this work we set $\alpha = 0.29$, which we obtained from a
least-squares fit of Eq.~\eqref{eq:ave_vf_3D} to the HS simulation
data for $\Avf$ versus $\eta$ over the range $0.375 \le \eta \le
0.525$.  The quality of this fit is illustrated in the inset of
Fig.~\ref{fig:HS-fv}a.

We also compare in Fig.~\ref{fig:HS-fv}a the free-volume distributions $f(\vf)$
obtained from simulations with those predicted via
Eq.~\eqref{eq:3Dify}.  We observe that the model describes the data
relatively well, with the predictions becoming semi-quantitative for
the higher packing fractions investigated.  This trend with packing
fraction is expected since the simulated free volumes become more
compact at higher densities, and, as a result, they more closely
resemble the simple cuboid shapes that we have assumed in our model.

The same free-volume distributions shown in
Fig.~\ref{fig:HS-fv}a are also shown in Fig.~\ref{fig:HS-fv}b, only now scaled in the manner suggested by
Eq.~\eqref{eq:3Dify_scaled}.  This rescaling yields the probability
density associated with observing a particle with a particular value
of $\vfs =\vf/\Avf$.  Interestingly, we observe that, to a very good
approximation, the scaled distributions obtained via simulation for
all packing fractions investigated collapse onto a single curve
predicted by the model.

\begin{figure}[t]
  \centering
  \includegraphics{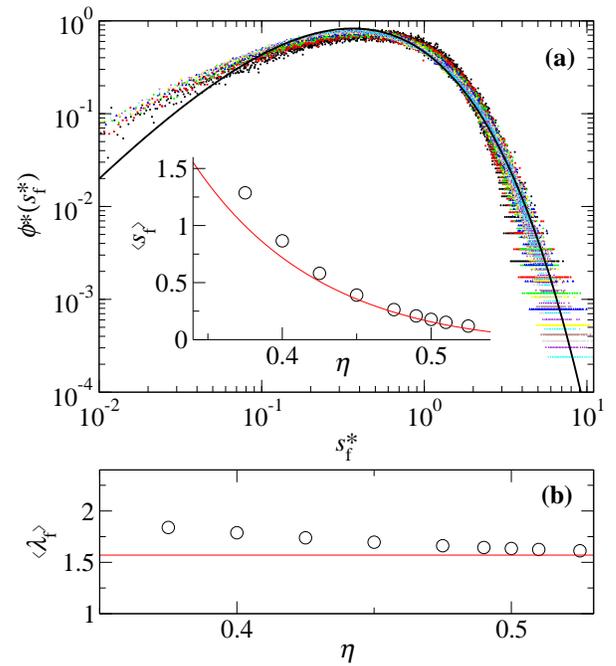}
  \caption{(a) Proabability density associated
    with observing particles with scaled free surface area
    $\sf^*=\sf/\Asf$ in the 3D HS fluid for
    packing fractions $\eta=0.375,\ 0.4,\ 0.425,\ 0.45,\ 0.475,\
    0.49,\ 0.5,\ 0.51$, and 0.525.  Symbols are simulation data and solid curves are
    the predictions of Eq.~\eqref{eq:3Dify_scaled_fs}.  Inset: Average free surface area
    $\Asf$ versus packing fraction $\eta$.  Circles are simulation
    data and the solid line is the prediction of
    Eq.~\eqref{eq:ave_sf_3D}.  (b)
    Average sphericity $\left< \lambda_{\mathrm{f}} \right>$ of free volumes in the HS fluid 
versus packing fraction 
$\eta$.  Circles are simulation data, and the
    solid line is the prediction of Eq.~\eqref{eq:ave_sphereicity}.}
  \label{fig:HS-fs}
\end{figure}

As can be seen by Eq.~\eqref{eq:3Dify_scaled_fs}, the model 
predicts that the distributions of free surfaces for the HS fluid
should also collapse onto a single curve when scaled in an analogous way.  
We compare in Fig.~\ref{fig:HS-fs}a the scaled free-surface 
distributions $\phi^*(\sf^*)$ obtained via simulation to the predictions of 
Eq.~\eqref{eq:3Dify_scaled_fs}.  Again, the collapse of the simulation
data is striking, and it is described reasonably well by the
free-volume model.  The inset of Fig.~\ref{fig:HS-fs}a
compares the average free-surface area per particle $\Asf$ obtained 
from simulations to the prediction of Eq.~\eqref{eq:ave_sf_3D}.
As with the predictions for $f(\vf)$, the model provides a more
accurate description of $\Asf$ at higher packing fractions where the 
free volumes are expected to be more compact.  
At low packing fractions, particles will on average have more nearest neighbors which define their free-volume then at high packing .  Therefore, we expect our cuboid approximation to break down at low packing fractions.

To quantify the compactness of the HS free volumes, 
we compare in Fig.~\ref{fig:HS-fs}b the average sphericity of the
free volumes $\left< \lambda_{\mathrm f} \right>$ 
calculated from the simulations to that predicted by 
Eq.~\eqref{eq:ave_sphereicity}.  
We find that while the cuboid representation of free volumes in our model 
underestimates $\left< \lambda_{\mathrm f} \right>$, it qualitatively captures 
the fact that, on average, the shapes of the free volumes in the HS fluid 
are fairly insensitive to changes in packing fraction.

Finally, we test the ability of our free-volume 
model to predict the equation of
state of the HS system.  In particular, we compare
Eq.~\eqref{eq:eos_sf_vf} to the well-known
Carnahan-Starling\cite{Carnahan1969Equation-of-Sta} equation, 
which provides an accurate description of the HS pressure for
packing fractions in the equilibrium fluid range $0 < \eta < 0.494$.
\begin{figure}
  \centering
  \includegraphics{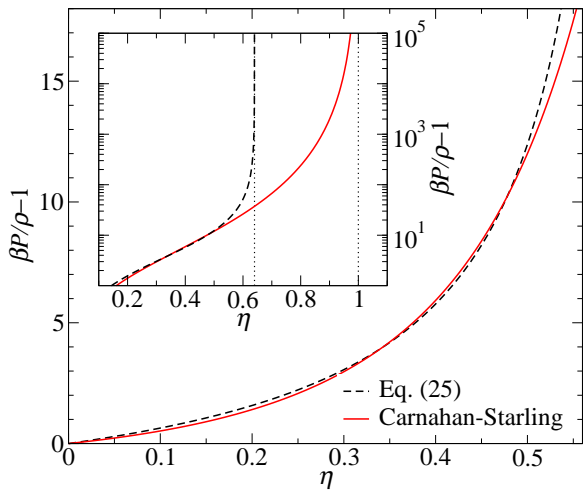}
  \caption{Equation of state for the HS fluid as calculated
    from both Eq.~\eqref{eq:eos_sf_vf} and the
    Carnahan-Starling\cite{Carnahan1969Equation-of-Sta} relationship.
    Inset: Expanded version illustrating that Eq.~\eqref{eq:eos_sf_vf}
    diverges at $\eta_{\mathrm{MRJ}}=0.64$, while the Carnahan-Starling
    relationship diverges at the unphysically high packing fraction of
    $\eta=1$.}
  \label{fig:eos}
\end{figure}
As can be seen in Fig.~\ref{fig:eos},
Eq.~\eqref{eq:eos_sf_vf} shows good agreement with
the Carnahan-Starling equation for the equilibrium HS fluid.  
Moreover, 
while the Carnahan-Starling equation
diverges at the unphysically high packing fraction of $\eta=1$, 
Eq.~\eqref{eq:eos_sf_vf} predicts that the HS fluid becomes
incompressible at the packing fraction of the maximally random jammed state 
($\eta_{\mathrm{MRJ}}=0.64$).\cite{Torquato2000}
\subsection{Predictions for the Square-Well Fluid} 
To demonstrate that the ideas outlined in Section II can 
be readily extended to other fluid systems, we now use our approach to 
predict the free-volume 
distributions of a square-well fluid with short-range attractions, 
a basic model system for suspensions
of attractive colloids and globular proteins. 
Fluids with short-ranged attraction are 
of particular interest here because they exhibit 
free-volume distributions with shapes that differ from that of
the equilbrium HS fluid at the same packing fraction.\cite{krekelberg2006}  
Specifically, interparticle 
clustering induced by the short-range attractions 
increases the populations of both large and small free volumes at the
expense of mid-sized free volumes.  These structural changes play an
important role in the anomalous dynamical properties exhibited by
these fluids,\cite{krekelberg2006} and the ability to predict such changes
constitutes a sensitive test of our free-volume model.

The square-well pair potential is given by
\begin{equation}
  \label{eq:vSWij}
  V_{ij}(r)=
  \begin{cases}
    \infty&r<1,\\
    -\epsilon&1\leq r < 1+\Delta,\\
    0&r\geq 1+\Delta.
  \end{cases}
\end{equation}
Using Eq.~\eqref{eq:vSWij}, \eqref{eq:eos1d-1}, and \eqref{eq:eos1d-2},
one can show that the equation of state of the 1D square-well fluid
can be expressed as
\begin{equation}
  \label{eq:eos_sw_1d}
  \xi^{-1}=1+\frac{1}{\beta \Pi}+\frac{\Delta e^{-\beta\Pi\Delta}(1-e^{-\beta\epsilon})}{1-e^{-\beta\Pi\Delta}(1-e^{-\beta\epsilon})}.
\end{equation}
Furthermore, by substituting Eq.~\eqref{eq:gap_def}, \eqref{eq:vSWij}, 
and~\eqref{eq:eos_sw_1d} into Eq.~\eqref{eq:pvf_1D_gen}, 
one finds that the corresponding 1D free-volume distribution is given by
\begin{equation}
  \label{eq:1Dfv_sw}
  \frac{\pvfo(\xf)}{\Theta(\xf)}=
  \begin{cases}
    \xf  & \xf<\Delta,\\
    2e^{-\beta\epsilon}(\xf-\Delta)-\xf+2\Delta & \xf \in
    [\Delta,2\Delta),\\ 
    \xf e^{-2\beta\epsilon} + 2\Delta e^{-\beta\epsilon} (1-e^{-\beta\epsilon}) & \xf\geq
    2\Delta,
  \end{cases}
\end{equation}
where
\begin{equation}
  \label{eq:prefac_sw}
  \Theta(\xf)=e^{-\beta \Pi \xf} \left[ \frac{\beta \Pi}{1-e^{-\beta \Pi \Delta}(1-e^{-\beta\epsilon})} \right]^2.
\end{equation}
By substituting Eq.~\eqref{eq:1Dfv_sw} into Eq.~\eqref{eq:3Dify}, one
can readily calculate the free-volume distribution for the 3D square-well
 fluid numerically.

\subsection{Simulations of the Square-Well Fluid}
To test the theoretical predictions of our model for the 3D 
square-well fluid,
we have performed a series of event-driven molecular dynamics
simulations\cite{Rapaport2004The-Art-of-Mole} using $N=1000$
particles.  As with the HS simulations described earlier, a cubic 
simulation cell was employed with periodic boundary conditions.  
Equilibrium configurations were 
stored during these runs and later 
used to
calculate\cite{Sastry1998Free-volume-in-,Sastry1997statistical-geo}
the geometric properties of the single-particle free volumes. For all simulations, 
the range of the square-well attraction
was set to $\Delta=0.03$.  
We studied the fluid at packing fraction $\eta=0.5$ and $0.58$. For the $\eta=0.58$ case, a flat distribution of particles sizes with mean $\sigma$ and half width $\delta=\sigma/10$ was used to prevent crystallization. We examined the behavior of the fluid for
various
values of the strength of the interparticle attraction~$\epsilon$,
which is often quantified using 
the reduced second virial coefficient
$B_2^*=B_2/B_{2}^{\mathrm{HS}}$. Here, $B_{2}=(1/2)\int d{\bf r}
  [1-\exp(-\beta V_{ij}(|{\bf r}|))]$ is the second virial
coefficient of the fluid of interest,
and $B_{2}^{\mathrm{HS}}=2\pi/3$ is
that of the hard-sphere fluid.  
For the 3D square-well fluid, $B_2^*$ is given by
\begin{equation}
  \label{eq:virial}
  \begin{split}
    B_2^*&=3\int_0^\infty r^2[1-e^{-\beta V_{ij}(r)} ]dr  \\
    &=1+(1-e^{\beta\epsilon})[(1+\Delta)^3-1].
  \end{split}
\end{equation}

\begin{figure}[t]
  \centering
  \includegraphics{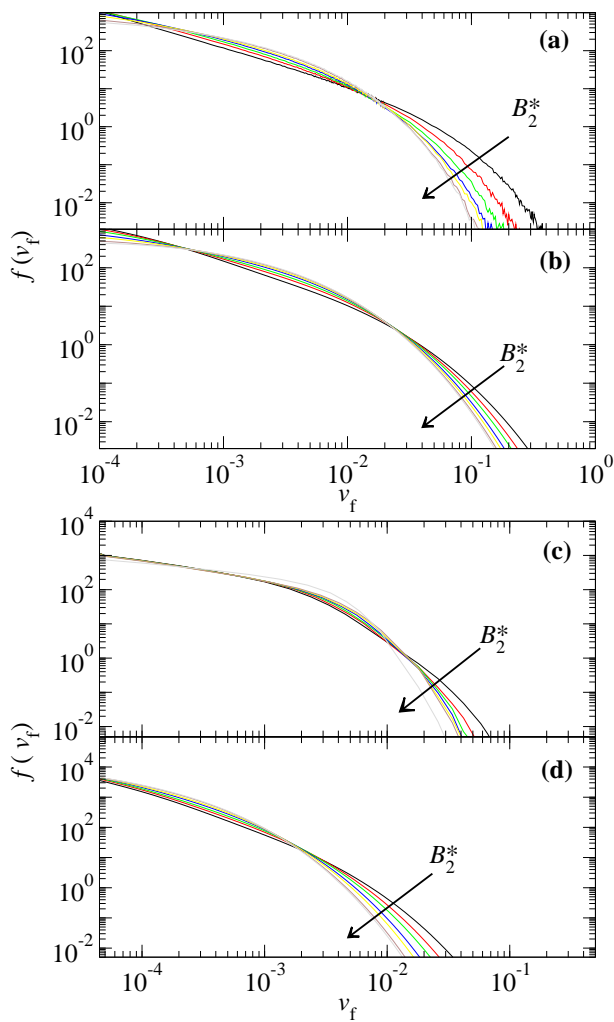}
  \caption{Free-volume distributions for the square-well fluid at
    range of attraction $\Delta=0.03$, and reduced second virial
    coefficients $B_2^*=-0.04,\ 0.41,\ 0.60,\ 0.78,\ 0.84,\ 0.91,$ and
    $0.94$. (a) Simulations at packing fraction $\eta=0.5$ and (b) the
    free-volume model at described in the text at $\eta=0.5$. (c)
    Simulations at packing fraction $\eta=0.58$ and (d) free-volume
    model at $\eta=0.58$.  Arrows indicate increasing $B_2^*$.  }
  \label{fig:swcomparison}
\end{figure}

The simulated results for the 
free-volume distributions are displayed in Fig.~\ref{fig:swcomparison}a and~\ref{fig:swcomparison}b for $\eta=0.5$ and $0.58$, respectively.  As has been observed for other 
model fluids with short-range attractions,\cite{krekelberg2006} decreasing
$B_2^*$ (increasing attractions) increases the populations
of both large and small free volumes at the expense of the mid-sized
free volumes.  These trends are captured by the predictions of the
free-volume model (Fig.~\ref{fig:swcomparison}b and~\ref{fig:swcomparison}d) 
using Eq.~\eqref{eq:3Dify}, \eqref{eq:1Dfv_sw},
and \eqref{eq:prefac_sw}.  It is evident that the model does not match the simulation results as well at $\eta=0.58$ as at $\eta=0.5$.  This is most likely due to the 1D equation of state used in the model being less accurate at high packing fractions.  Although the free-volume
model presented here does not reproduce the square-well 
distributions with quanitative accuracy, it clearly captures the important
physical trends. Moreover, its predictions show surprisingly good agreement
with the simulation data, given that there are no free 
parameters in the theory that are fit to data for the square-well fluid. 
\section{Conclusions}
In summary, we have introduced a simple model for predicting the 
free-volume distributions of equilibrium fluids, and we 
have tested this model using molecular simulations.  
The model suggests a new scaling for the
density-dependencies of the free-volume and free-surface 
distributions of the HS
fluid, and these scalings show very good agreement with
simulation results.  The free-volume model 
also predicts a reasonably accurate hard-sphere equation of state in
which the pressure
diverges at a packing fraction of $\eta=0.64$.  Finally, the model 
predicts semi-quantitatively the manner in which the attractive strength
affects the free-volume distributions of a square-well fluid with short-range 
attractions.  These considerations suggest that the free-volume model
proposed here can perhaps be used fruitfully within the free-volume
based theories for dynamics of fluids 
to understand how their transport coefficients derive from
their microscopic interactions and thermodynamic conditions.
\vspace{2cm}

{\textbf{Acknowledgments.}  WPK acknowledges financial support of the
National Science Foundation for a Graduate Research Fellowship.  TMT
acknowledges financial support of the National Science Foundation (CTS
0448721) and the David and Lucile Packard Foundation, and VG
acknowledges financial support of the Robert A. Welch Foundation and
the Alfred P.~Sloan Foundation.  Simulations were performed at the
Texas Advanced Computing Center (TACC).}

\end{document}